\documentclass[letterpaper]{article}

\usepackage{natbib,alifeconf}  

\title{A multiscale model of urban morphogenesis}
\author{Juste Raimbault$^{1,2,3}$\\
\mbox{}\\
$^1$Center for Advanced Spatial Analysis, University College London, London, UK\\
$^2$UPS CNRS 3611 ISC-PIF, Paris, France\\
$^3$UMR CNRS 8504 G{\'e}ographie-cit{\'e}s, Paris, France\\
juste.raimbault@polytechnique.edu} 

\begin{document}
\maketitle

\begin{abstract}
The dynamics and processes of urban morphogenesis are a central issue regarding a long-term sustainability of urban systems. They however imply stakeholders and parameters at multiple temporal and spatial scales simultaneously, leading to intricate interactions between dimensions and scales. We introduce in this paper from a theoretical viewpoint a simple agent-based model of urban morphogenesis at the scale of an urban area, with the feature of integrating the microscopic and the mesoscopic scales. At the local level, developer agents drive urban development conditioned by local properties but also an infrastructure network at a smaller scale. The network evolves more slowly following global properties. Indicators of sustainability including modal shares and urban density suggest an application of the model to multi-objective optimisation. We finally discuss possible implementation, extensions and applications of the model.
\end{abstract}

\section{Introduction}

Artificial life approaches to the modeling of complex systems are by essence interdisciplinary, and have been proved relevant in many disciplines of social science \citep{youn2018scaling}. In particular, urban science - understood in a broad way as a set of quantitative attempts to model and simulate processes of urbanisation across scales and time periods \citep{batty2013new} - has a long tradition of using such methodologies (e.g. agent-based models or cellular automata models of urban growth) or concepts imported from life science such as urban ecology or morphogenesis \citep{raimbault2020cities}. Biological metaphors have also historically been used in urban planning \citep{batty2009centenary}.

Understanding processes occurring within and between cities, and the structures emerging from these at different scales, is a challenge more relevant than ever in the context of global change and deprived urban conditions. Cities could indeed be as much a solution to sustainability issues than the endogenous source of many problems, by adapting to climate change and driving innovation \citep{pumain2020theories}. In this field, numerous complementary dimensions of urban systems can be studied and acted upon through planning and policies. For example, urban morphology can be quantified using fractal methods \citep{chen2010modeling}. The management of urban traffic directly impacts mobility patterns and emissions \citep{yoshioka2017macroscopic}. More generally, the use of ``living technology'', in the sense of importing concepts and models from Artificial Life, has been suggested by \cite{gershenson2013living} as a path towards more liveable cities. Cities can be simulated themselves as learning agents as \cite{goldstein2000cities} puts it for the case of co-evolving cities and wildfires.

The modeling of urban morphogenesis, in the sense of the dynamics of urban form and its link with urban function, enters this context and is particularly relevant regarding planning sustainability on the long term. \cite{raimbault2019generating} have explored complementary generative models of urban form at a very detailed scale. \cite{raimbault2018multi} show that urban road networks topologies correspond to different growth heuristics. \cite{raimbault2018calibration} shows that a simple reaction-diffusion model of urban growth including population only can already reproduce a broad set of existing urban forms. At the scale of the urban system, the concept of evolution and co-evolution within systems of cities is more appropriate than morphogenesis to describe urban dynamics \citep{pumain2018evolutionary}. \cite{wu2011urban} propose to use calibrated parameters of urban growth as an urban DNA driving urban evolution. \cite{silver2019towards} propose an extension of social evolution based on a refinement of the concept of meme, assuming elementary urban patterns that can be transmitted, mutated and selected. \cite{raimbault2020model} develops a model of urban evolution based on the diffusion of innovation, in which the urban genome is constituted by adoption proportions within each city.

Among these efforts to model urban dynamics processes, proper multiscale models remain less explored. Artificial life approaches include tackling multiscale problems and to quantify interactions between scales \citep{seth2010measuring}. Regarding urban dynamics, some prior work focus simultaneously on different scales. \cite{batty2005agents} introduces a common formalisation using cellular automata to simulate urban dynamics from the neighbourhood scale to the system of cities scale. \cite{cheng2003modelling} use logistic regressions to investigate drivers of urban growth at multiple scales, distinguishing between probability of urbanisation at the urban area level and density of change at the district level. \cite{zheng2017decision} couple a system dynamics model at the city scale with a land-use change model at the local scale to evaluate the sustainability of urban renewal initiatives. \cite{torrens2012polyspatial} introduce a framework with household agents relocating across scales. A similar approach is adopted by \cite{murcio2015urban} with the simulation of migrations at different spatial ranges. Issues related to urban energy consumption and urban climate have also extensively been studied from a multi-scale perspective. \cite{lim2017multi} couple climate models with local urban microclimate models to allow accounting for the urban heat island effect in planning. In a similar context, \cite{wong2021integrated} introduce a strong coupling between the scale of the street and the scale of the building.

This literature however does not directly address the question of urban morphogenesis,  and generally does not simulate a strong coupling between scales implying upward and downward causation. This paper introduces a new model for urban morphogenesis, with the particular feature of coupling dynamics at the microscopic scale (building, neighborhood and developer agents) with dynamics at the mesoscopic scale (transportation network). This model builds on \citep{raimbault2014hybrid} for its structure and land-use dynamics, and on \citep{le2015modeling} for the transport network evolution. Our contribution relies mostly on (i) a stylised and simple model which can be systematically explored; (ii) a strong integration between the micro and meso scales. We remain in this work at the level of a theoretical description of the model, suggesting paths towards implementation, developments and applications.

The rest of the paper is organised as follows: we first develop the rationale for the multi-scale model, and then describe it formally. We provide a detailed description of model parameters and indicators to quantify the sustainability of generated urban forms. We finally discuss possible model applications and extensions.

\section{Multi-scale urban morphogenesis model}

\subsection{Rationale}

It is a well-known stylised fact in urban systems that the land-use system and the transport system, at least its transport infrastructure component, evolve at different temporal scales. This assumption is fundamental in the development of land-use transport models which simulate the evolution of land-use following a controlled modification of accessibility through a new transport project \citep{wegener2004overview}. These models have been extended to long spatial ranges, such as in the case of the Quant model which is applied at the scale of a whole country \citep{batty2021new}. They however do not consider long temporal scales, at which infrastructure networks evolve \citep{lagesse2015spatial}. To consider both aspects at the same time, \cite{raimbault2018caracterisation} has introduced and studied the concept of a co-evolution between transportation networks and territories. Our model directly enters this theoretical framework. Considering the mesoscopic spatial scale of an urban area, we consider on the one hand a public transport network, evolving relatively slowly following some governance decisions, and on the other hand the dynamics of land-use with buildings being built relatively fast. The land-use model extends the rationale of \cite{raimbault2014hybrid} by introducing developer agents with a stochastic discrete choice behavior instead of a deterministic development of cells with a higher utility. This development is driven by essential explicative variables, which relative influence allows controlling the morphology of the emerging city. These variables are conditioned by the transport network, as distance to stations and distance to urban centres, and also depend on the local urban context as densities. We consider two types of land-use, namely housing and offices. The transport network evolution rules build on \cite{le2015modeling} for its formalisation as a two mode network in a zoned urban area, and on \cite{raimbault2018multi} for the different network growth heuristics included.

\subsection{Model description}

The urban morphogenesis model operates at the scale of an urban area, typically between 50 and 100km in practice. The urban space is described at the mesoscopic level by a grid of patches, which we take to simplify as a square world of width $W$. A transportation network is overlaid over these patches: potential nodes of the transportation network are patch centres. At the microscopic scale, the urban space consists in a vector description of buildings as polygons. Land-use is described in a stylised way through a binary typing for buildings, which can be either housing or office space. Buildings have furthermore a number of storeys, which is an important variable to determine local density. The transportation network includes a background grid linking all neighbour patches (corresponding to a diffuse road network), on top of which public transport nodes and links with a significantly higher speed (relative speed parameter $v_0$) are built. Some privileged nodes of the public transport network are coined as urban centres at the initial step. These centralities are assumed to include essential amenities and the distance to them will be an explicative variable to determine urban growth.

\subsubsection{Building dynamics at the microscopic scale}

The dynamics at the microscopic scale is driven by the development of new buildings, through the intermediary of $D$ developers agents. At each time step, each developer will achieve a single construction project, which composed of $f$ units of floor space localised in a given neighborhood. They have for this project some choice to make on (i) the share for each type of floor space (dwellings or offices); (ii) the location of the new buildings; and (iii) the density level. In the original model developed by \cite{raimbault2014hybrid}, location choice is deterministic and the patches with the highest utility value are built. We generalise the model by using a discrete choice model for location and land-use shares, considering the utility

\[
U_{i,k,d} = s_k \cdot \sum_j w_{j,d}^{(h)} d_j^{(h)}(i) + (1 - s_k) \cdot \sum_j w_{j,d}^{(w)} d_j^{(w)}(i)
\]

where $i$ stands for patch number $i$, $s_k \in \left[0 ; 1\right]$ the $k$-th level for the share of dwellings (in practice, a discrete number of shares are tested), $d_j^{(h)}(i)$ (resp. $d_j^{(w)}(i)$) is the $j$-th explicative variable for home land-use and patch $i$ (resp. for work land-use), and $w_{j,d}^{(h)}$ (resp. $w_{j,d}^{(w)}$) are the utility weights of different explicative variables for developer $d$ for home land-use (resp. work land-use). In practice, we follow the specification of \cite{raimbault2014hybrid} and consider a reduced number of explicative variables, which are: (i) $d_0 = d_R$ the distance to the transportation network; (ii) $d_1 = d_C$ the distance to the closest center; and (iii) $d_2 = \rho$ the home or work density around the patch. All explicative variables are at each time step rescaled across patches between 0 and 1 to compute utilities, such that weight parameters are also scaled such that $w_{j,d} \in \left[-1 ; 1\right]$.

This density explicative variable is computed as a spatial weighted average around the patch, using an exponential kernel of width $\lambda$. More precisely, the average is given for population by

\[
\rho (\vec{x}_i) = \frac{\lambda}{2\pi} \cdot \int_{r=0}^{\infty} \int_{\theta=0}^{2\pi} p(\vec{x}_i + r\cdot \vec{u}_{\theta}) \cdot \exp\left(- \lambda \cdot r\right) dr d\theta
\]

and by a similar expression for employments.

At the first order, accessibility to population or to employments \citep{levinson2020towards} may be approximated by local density if the range considered is of the same order than $\lambda$, and we thus do not consider such additional explicative variables. Note that additional dimensions of land-use in the model, such as green space, would imply to consider related explicative variables such as accessibility to green spaces \citep{higgs2012investigating}.

For each developer agent, utilities for all patches and all shares are computed, and the choice is drawn following the discrete choice probability given by

\[
p_{i,k,d} = \frac{e^{\beta \cdot U_{i,k,d}}}{\sum_{i,k} e^{\beta \cdot U_{i,k,d}}}
\]

where $\beta$ is a model parameter capturing the randomness of choices.

The agent then makes the choice of the density level of the building project. We assume at this stage that projects in a neighborhood will tend to be similar to what already exists, and thus the density is drawn following a normal law with average the local density $\rho = \frac{\rho^{(h)} + \rho^{(w)}}{2}$ and standard deviation $\sigma_d$ a parameter of the agent (propensity to take risk by investing in projects changing the shape of neighbourhoods).

To simplify, we consider only square buildings, of width randomly drawn for each building within reasonable boundaries (between 10m and 50m). Buildings can not overlap and must be spaced by a fixed street buffer (that we take as 10m in practice). We do not consider elaborated block filling algorithms such as the ones developed by \cite{brasebin2017apports}, but consider that large buildings correspond to such building aggregations. Iteratively, the agent (i) draws a building size; (ii) computes the corresponding number of floors required to reach the chosen density; (iii) adds the building at coordinates satisfying the previous constraints, within the chosen patch and minimising the distance to the closest station or centre; (iv) updates its floor space remaining to be built and iterates this procedure until the full project has been completed.

\paragraph{Transportation network dynamics at the mesoscopic scale}

Transport network is evolved each $t_m$ time steps, fixing the mesoscopic time scale. A fixed quantity of infrastructure length $I$ is constructed at these steps. First, if some patches are above a fixed distance threshold $\theta_d$ from the infrastructure network and above an activity threshold such that $p_i + e_i > \theta_a$, these are connected to the network by adding a link connecting to the closest link (either with an orthogonal projection and a new station or to the closest station depending on the configuration). These patches are connected in an iterative way, selected by decreasing distance to the network, such that network is built iteratively and no redundant links are created. This procedure is run as long as there remains some network length to be built or all such patches are connected.

Potentially new stations are open on the network, to simulate a policy intervention to tackle the so-called ``tunnel effect'' \citep{gonzalez2019long}, corresponding to a situation where territories are not served by a high-speed transport infrastructure traversing them. Considering patches above the $\theta_a$ threshold in terms of activity, and below $\theta_d$ in terms of distance to the network but above it in terms of distance to the closest station, a new station node is added on the closest link at the location of the orthogonal projection of the patch centroid.

Finally, potentially long-distance, redundant or looping links are added, as transport projects answering to some demand through governance processes \citep{le2015modeling}. Empirical evidence for processes driving such transport projects are diverse, and no single mechanisms has been exhibited to explain growth dynamics of public transport network, as \cite{raimbault2018multi} shows that multiple mechanisms are complementary to generate existing topologies for road networks. We follow this multi-modeling paradigm, and given a governance scenario set by a vector parameter of probabilities $\vec{n}_g$ for each governance processes, one rule is uniformly drawn among these processes and a link is added following this rule. In practice, we consider the heuristics described in \cite{raimbault2018multi}: a deterministic gravity potential breakdown, a random potential breakdown, a cost-benefit rule, and a biological network generation method. This step is potentially iterated until all infrastructure length $I$ has been built.

\paragraph{Model dynamics: strong coupling the micro and meso scales}

The model proceeds iteratively from an initial state, through the following stages at each mesoscopic time step:
\begin{enumerate}
	\item Network distances are updated, to compute patch distances to stations and to urban centres, while weighted spatial averages are computed for local densities.
	\item Given these explicative variables, the microscopic step where developers add buildings is iterated $t_m$ times, with density variables being updated at each time step.
	\item Population and employment patch aggregates are recomputed, to determine gravity potentials used in network growth heuristics.
	\item Network is evolved following the network growth procedure described above.
\end{enumerate}

The model is stopped until a total number of time step has been reached, or when a total population or employment criteria has been reached.

\subsection{Model setup}

The model can in principle be initialised with any realistic configuration, given some building and public transport network data. As some open dataset such as OpenStreetMap have a relatively good quality regarding these dimension \citep{arsanjani2015quality}, at least in Europe, this can be reasonably considered. However, given the stylised nature of the model, we propose an initialisation on synthetic urban systems. This approach furthermore allows designing numerical experiments to discriminate between intrinsic model dynamics and the contingent effects of geography \citep{raimbault2019space}.

We consider a setup similar to the model in \cite{raimbault2014hybrid}: a small number of initial centres, linked by a backbone transport network without any loop and obtained with a connection algorithm. For each of these centres, we setup a fixed number of identical buildings (same width and height, and thus same density), with a fixed initial share between housing and offices (in practice taken as 0.5).


\subsection{Model indicators}

We consider urban indicators among different dimensions. We quantify final configuration using:

\begin{itemize}
	\item urban morphology indicators, namely global density, spatial autocorrelation and average distance between individuals \citep{raimbault2018calibration} - these allow in particular characterising sprawled configuration;
	\item indicators on how land-use is mixed, using dissimilarity indicators on patches;
	\item network performance indicators, including average speed and the number of detours;
	\item sustainability indicators: we estimate a public transport mode share by assuming a probability to take public transport inversely proportional to the distance to the closest station, and integrate this share on the full population - the maximisation of this share will decrease emission linked to transport; we also estimate the average commuting trip distance between all population and all employments, which minimisation will imply a more local commuting and also less emissions.
\end{itemize}








\section{Discussion}

\subsection{Implementation and applications}

We have only given a theoretical and formal description of our multiscale model of urban mmorphogenesis. We suggest that a NetLogo implementation would be a relevant approach to easily visualise generated urban forms, while a direct integration into the OpenMOLE software \citep{reuillon2013openmole} with for example a scale implementation would directly allow the application of advanced model exploration and validation methods.

Aspects that can be immediately explored include: (i) an exploration of the diversity of urban morphologies generated, in terms of morphology indicators, of types of transportation networks, but also in terms of interactions between the two - such a diversity can be obtained for example with the application of a diversity search algorithm such as the Pattern Space Exploration algorithm \citep{cherel2015beyond}; (ii) an exploration of the link between model parameters (in particular the utility weights $w_{j,d}$ which can be influenced by policies such as transit-oriented development or high density policies), generated urban morphologies and their sustainability - in particular the application of multi-objective optimisation should allow establishing if objective are contradictory and exhibit a compromise Pareto front; (iii) quantify the interactions between scales, in particular if the model exhibits upward and/or downwards causation, applying for example the method introduced by \cite{seth2010measuring} - this last application has theoretical implications regarding the multiscale modeling of complex systems, and of urban systems in particular.

\subsection{Further developments}




We have described from a theoretical and formal viewpoint a relatively simple model, which remains only loosely linked to empirical models of urban systems through stylised facts and assumptions. A data-driven approach with a more realistic parametrisation for utility formulation, explicative variables and weights, would be closer to cellular automaton approaches to urban modeling \citep{clarke2007decade} and would imply a more explicit bridge to planning and policies. Similarly, investigating the link with existing literature in urban and regional modeling, in which land and housing markets are explicitly included (see for example the Nedum-2D model which remain empirical and data-driven \citep{viguie2012trade}), would provide a better contextualisation of results obtained through model exploration.

As already stated, the inclusion of additional dimensions and explicative variables, such as the accessibility to green space, or different types of amenities, or more generally any type of accessibility, is also an important direction to explore. In particular, to what extent the diversity of urban morphologies is linked to the variety of underlying explicative factors remains an open question. The application and calibration of the model on real systems is also a relevant research direction towards the study of sustainable policies. Finally, more theoretical considerations can be suggested, such as the integration of this model into a meso-macro model at the scale of the system of cities, which would set the basis for a multiscale model with a strong integration between three distinct scales and corresponding processes \citep{rozenblat2018conclusion}. Similarly, a non-stationary version of this model, where agents have parameters depending on regional characteristics, would provide relevant insights into the spatio-temporal non-stationarity of territorial systems.

\section{Conclusion}

We have given a theoretical and formal description of a multiscale model of urban morphogenesis. This model couples for one urban area the fast evolution of land-use through developer agents with the slow evolution of an infrastructure network through governance processes. In that sense, the model is furthermore a co-evolution model between transportation networks and territories. We have suggested directions for implementation and applications, including the exploration of sustainable policies to design and plan cities while acknowledging some part of their complexity.

%

\footnotesize

\end{document}